\documentclass[twocolumn,aps,pra]{revtex4}
\usepackage{epsfig}
\usepackage{amsmath}
\usepackage{amsfonts}
\usepackage{color}
\usepackage{soul}
\def\a{\hat{a}}
\def\M{\hat{M}}
\def\N{\hat{N}}
\def\F{\hat{F}}
\def\sig{\hat{\sigma}}
\def\NN{\mathcal{N}} 
\def\MM{\mathcal{M}} 	
\def\AA{\mathcal{A}} 
\def\FF{\mathcal{F}}
\def\DD{\mathcal{D}}
\def\AAA{\mathbb{A}}
\def\FFF{\mathbb{F}}
\def\DDD{\mathbb{D}}
\def\XXX{\mathbb{X}}

\begin{document}

\title{Active optical frequency standard using sequential coupling of atomic ensembles}
%\title{Prospect for optical lattice active frequency standard with optical conveyor}
\author{G.~A.~Kazakov$^{1,2,}$\footnote{kazakov.george@gmail.com}, T.~Schumm$^1$}
\affiliation{{\setlength{\baselineskip}{18pt}
$^1${Vienna University of Technology, Atominstitut, Stadionallee 2, 1020 Vienna, Austria}\\
$^2${St. Petersburg State Polytechnic University, Polytechnicheskaya 29, 195251 St. Petersburg, Russia}
}}

%-------------------------------abstract-------------------------------------------
\begin{abstract}
Recently, several theoretical proposals addressed the generation of an active optical frequency standard based on atomic ensembles trapped in an optical lattice potential inside an optical resonator. Using atoms with a narrow linewidth transition and population inversion together with a ``bad'' cavity allows us to realize the superradiant photon emission regime. These schemes reduce the influence of mechanical or thermal vibrations of the cavity mirrors on the emitted optical frequency, overcoming current limitation in passive optical standards. The coherence time of the emitted light is ultimately limited by the lifetime of the atoms in the optical lattice potential. Therefore these schemes would produce one light pulse per atomic ensemble without a phase relation between pulses. Here we study how phase coherence between pulses can be maintained by using several inverted atomic ensembles, introduced into the cavity sequentially by means of a transport mechanism. We simulate the light emission process using the Heisenberg-Langevin approach and study the frequency noise of the intracavity field.

\vspace*{10pt}

\end{abstract}
\maketitle 

\allowdisplaybreaks
%-------------------------------------------------Introduction--------------------------------------------------------------
\section{Introduction}
\label{sec:intro}

The concept of an active frequency standard in the optical domain was recently proposed and studied by several authors~\cite{Chen05, Yu08, Chen09, Meiser09, Meiser10, Yang10, Zhuang11, Zhang12}. The main idea is to create a ``superradiant'' laser operating deep in the bad cavity regime where the decay rate $\kappa/2$ of the cavity field significantly exceeds the linewidth $\gamma_{ab}$ of the lasing transition. Then the fundamental linewidth $\Delta \omega_{ST}$ of the steady-state operating laser can be described by the generalized Schawlow-Townes formula that can be written as
\begin{equation}
\Delta \omega_{ST}=\frac{g^2 \NN_{a0}}{I_0 \gamma_{ab}} \left(\frac{\gamma_{ab}}{\kappa/2+\gamma_{ab}} \right)^2
\label{eq:1}
\end{equation}
in the practically interesting range $\Delta \ll (\gamma_{ab}+\kappa/2)$~\cite{Kolobov93, Kuppens94, Davidovich96}. Here  $\Delta=\omega_a-\omega_c$ is the difference between the cavity resonance frequency $\omega_c$ and the frequency $\omega_a$ of the lasing transition, $g$ is the coupling coefficient, $\NN_{a0}$ is the steady-state occupation of the upper lasing level, $I_0$ is an average number of photons in the cavity mode. The frequency $\omega$ of the emitted radiation is connected with the resonance frequency $\omega_c$ of the cavity and with the frequency $\omega_{a}$ of the atomic transition via cavity pulling, namely 
\begin{equation}
\omega = \frac{2 \omega_c \gamma_{ab}+\omega_{ab}\kappa}{2 \gamma_{a}+\kappa}.
\end{equation}
In the bad cavity regime, any fluctuation (thermal or mechanical) of the cavity length has much less influence on the spectrum of the output light than in conventional lasers, operating in the good cavity regime. Fluctuations of the cavity length are currently the main limiting factor for the performance and even more for the transportability of modern passive optical frequency standards. An active optical frequency standard could help to overcome these limitations.

Two main approaches towards an active optical frequency standard are proposed today: the {\em optical lattice laser}~\cite{Chen05, Meiser09, Meiser10} and the {\em atomic beam laser}~\cite{Yu08, Chen09, Yang10, Zhuang11, Zhang12}. The first approach suggests to use  trapped atoms with narrow optical transitions [such as  $^1S_0\leftrightarrow\, ^3P_i, \, (i=1,2,3)$ transitions in divalent atoms] confined to the Lamb-Dicke regime inside an optical lattice potential as a gain medium to build the laser. The necessary population inversion can be provided by additional repumping fields, coupling the lower lasing state with some higher levels from which the atoms decay to the upper lasing level (three-level laser scheme). The general operation principle of such an optical lattice bad cavity laser has been demonstrated experimentally using a Raman system to mimic a narrow linewidth optical transition~\cite{Bohnet12}. 

The main limitation to the performance of this optical lattice laser is the limited lifetime of the atoms in the optical trap. A laser whose gain medium is a single ensemble of trapped atoms can not keep the phase longer than the atom  trap lifetime. Another problem emerges from the pumping laser which can shift the frequency of the lasing transition. Although the authors of~\cite{Meiser09} claim that ac Stark shift-induced fluctuations of the atomic transition frequency are negligible at the mHz level for repumping rates of the order of $10^3$ s$^{-1}$ and for a  pumping laser intensity-stabilized to 1 \%, also the frequency of the laser depopulating the low lasing level needs to be quite well stabilized. In \cite{Zhang12}, it is proposed to use a four-level lasing scheme with a non stable lower lasing state to avoid the pumping-induced shift, however this scheme still suffers from the problem of limited trap lifetime.

The atomic beam approach allows us to overcome the atom lifetime limitation. Here, a continuous beam of active atoms, previously pumped to the upper lasing state, passes through the cavity. The theory of laser generation for this system was presented in~\cite{Yu08} and the generalization to Ramsey-type interaction between the atoms and the intracavity light was performed in~\cite{Yang10}. In~\cite{Chen09} the author claimed that laser radiation with $0.5$\,Hz linewidth and 120\,nW power can be attained with a beam of hot $^{88}$Sr atoms prepumped to the $^3P_1$ upper lasing state (natural linewidth 7.6\,kHz), with a cavity of 800\,$\mu$m waist and  $\kappa=11$\,MHz. The flux required for the lasing is $4.3\times 10^{11}$ atoms per second. At the same time, the theory developed in~\cite{Yu08} does not take into account the motion of the atoms along the cavity axis while passing through the resonator. This motion will lead to additional line broadening and a first-order Doppler shift. The latter will fluctuate together with the mean transversal atomic velocity which depends on the environment, thermal fluctuation of the nozzle etc. Accurate compensation of all these effects appears to be a very challenging task.  In~\cite{Zhuang11}, a similar idea is proposed for the creation of an extreme ultraviolet laser using metasable noble gas beams. 

We note that for lasing transitions with longer wavelengths (such as infrared transitions in some molecules) it may be possible to combine the Ramsey-type lasing scheme~\cite{Yang10} with mechanical blocking of the particles that would pass through the interaction regions with different phases of the field, similarly to~\cite{Kramer78}.

In the present work we investigate an intermediate, alternative approach to the schemes presented above, consisting in dynamically moving trapped atomic ensembles into (and out of) the cavity. These ensembles will be prepared in the upper lasing state outside the cavity, circumventing perturbations due to ac Stark shifts. Transport will be realized by means of a moving red-detuned one-dimensional (1D) optical lattice, transport of more than $2\times 10^{5}$ ultracold atoms over a distance of up to 20 cm with fast transport velocities of up to $6 \, \mathrm{m/s}$ and strong accelerations of up to $2.6 \times 10^3  \, \mathrm{m/s^2}$ have already been demonstrated experimentally~\cite{Schmid06}. To confine the atoms to the Lamb-Dicke regime inside the cavity, an auxilary blue-detuned stationary optical lattice is used in our scheme. Both, the red-detuned and blue-detuned optical lattice should have ``magic frequencies''~\cite{Derevianko11} providing equal light shift for both states of the lasing transitions.

\begin{figure}
\begin{center}
\resizebox{0.48\textwidth}{!}
{\includegraphics{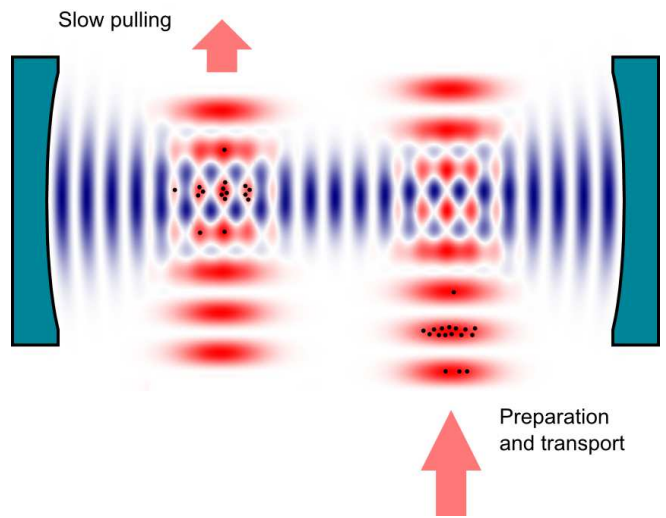}}

\end{center}
\caption{(Color online) Schematic of two atomic ensembles (black dots) in two moving optical lattices during the first stage of the lasing process. The first atomic ensemble emits light into the resonator mode while being slowly pulled through the cavity waist; the second one is prepared and transported towards the cavity.}
\label{fig:f1}
\end{figure}

The operational sequence for two ensembles consists of four stages.  In the first stage, one atomic ensemble is positioned inside the cavity and starts to emit light into the cavity mode, whereas a second ensemble is trapped outside the cavity, pumped into the upper lasing state, and transported towards the cavity; see Fig.~\ref{fig:f1}. In the second stage, when a significant part (but not all) of the atoms in the first ensemble are transferred to the ground state, the second ensemble is introduced into the cavity. The atoms of the second ensemble start to emit photons into the cavity mode mainly via stimulation emission, maintaining the optical phase. In the third stage the first ensemble is extracted from the cavity and a new inverted ensemble is prepared (either by loading new atoms or by repumping) while the second ensemble emits the light. The fourth stage is identical to the second one up to permutation of the ensembles. To avoid the time-dependent second-order Doppler effect, one should keep a constant velocity of the moving optical lattice while atoms cross the cavity waist. This process can be repeated as many times as necessary to keep the phase of the laser field. 

In the following, we model the lasing process using the Heisenberg-Langevin approach and study the phase stability of the emitted light.
%---------------------------------begin figure scheme crystal---------------------------
%---------------------------------end figure figure Th-doped crystal-----------------------------
%-------------------------------------------------Model--------------------------------------------------------------
\section{Model}
\label{sec:model}
%
%-----------Operator Heisenberg-Langevin equations--------------------------
\subsection{Operator Heisenberg-Langevin equations}
\label{sec:model_operator}
 We employ a model based on the set of quantum Langevin equations following~\cite{Kolobov93, Benkert90}. One or two ensembles $\AAA_1$ and $\AAA_2$ of homogeneously broadened two-level atoms with transition frequency $\omega_{ab}$ are held in a cavity characterized by a cavity damping constant $\kappa$. The finesse of the cavity is assumed to be sufficiently high to justify the mean-field approximation. The atoms interact with the radiation field of a single excited mode of the cavity which we consider as a plane wave with frequency $\omega_c$. By adding up the individual atomic operators, we define the macroscopic atomic operators
\begin{eqnarray}
\M^\chi(t) & = & -i\sum_{j \in \AAA_\chi} \sig^j_{ba}(t), \nonumber \\
\N_a^\chi(t) & = & \sum_{j \in \AAA_\chi} \sig^j_{aa}(t), \label{eq:3} \\
\N_b^\chi(t) & = & \sum_{j \in \AAA_\chi} \sig^j_{bb}(t), \nonumber
\end{eqnarray}
where $\chi=1,2$ is the index of the corresponding atomic ensemble, $\sig_{ba}^j=(|b\rangle\langle a|)^j$, $\sig_{aa}^j=(|a\rangle\langle a|)^j$, $\sig_{bb}^j=(|b\rangle\langle b|)^j$, $|a\rangle$ and $|b\rangle$ are the upper and lower lasing states, respectively. The equation for the photon annihilation operator $\a$ is
\begin{equation}
\dot{\a}=- 
\frac{\kappa}{2}\a+ g \left(\M^1 \Gamma_1(t)+\M^2 \Gamma_2(t) \right) +\F_\gamma(t), 
\label{eq:4}
\end{equation}
where $\Gamma_\chi(t)$ describes the time dependence of the coupling of the $\chi$th atomic ensemble with the cavity mode, $\F_\gamma$ is the Langevin force. Here we suppose for the sake of simplicity that all the atoms have the same coupling coefficient $g$ and that $\omega_{ab}=\omega_c$. The last approximation is valid if $|\omega_{ab}-\omega_c| \ll (\kappa/2+\gamma_{ab})$.

In our model we suppose that when a certain atomic ensemble starts to interact with the cavity field, all the atoms of this ensemble have been pumped into the state $|a\rangle$. During the interaction time, the set of equations for the macroscopic atomic operators (\ref{eq:3}) of the $\chi$th ensemble is
\begin{eqnarray}
\dot{\M}^\chi&=&-\gamma_{ab} \M^\chi + g \Gamma_\chi(t) \left(\N^\chi_a-\N^\chi_b \right) \a +\F_M^\chi(t), \nonumber  \\
\dot{\N}_a^\chi&=&-\gamma_{a} \N^\chi_a - g \Gamma_\chi(t) 
\left(\a^+\M^\chi+\M^{\chi +} \a \right) +\F_a^\chi(t), \nonumber \\
\dot{\N}_b^\chi&=&\gamma_{a} \N^\chi_a + g \Gamma_\chi(t) 
\left(\a^+\M^\chi+\M^{\chi +} \a \right) +\F_b^\chi(t), \label{eq:5}
\end{eqnarray}
where $\gamma_a$ is the spontaneous decay rate of the upper lasing state, $\gamma_{ab}$ is the decay rate of atomic coherence, and $\F_M$, $\F_a$, and $\F_b$ are the Langevin forces. Here we neglect loss of atoms from the trap during the interaction time. It is easy to show that $\F_a=-\F_b$ in this case. The mean values of the Langevin forces are~\cite{Kolobov93}
\begin{equation}
\left\langle \F_a^\chi(t)\right\rangle=\left\langle \F_b^\chi(t)\right\rangle=\left\langle \F_M^\chi(t)\right\rangle=\left\langle \F_\gamma(t)\right\rangle=0. \label{eq:6}
\end{equation}
Correlation functions of the Langevin forces corresponding to the intracavity field are
\begin{align}
&
\left\langle \F_\gamma^+(t) \F_\gamma(t') \right\rangle = \kappa \bar{n}_T \delta(t-t'), \label{eq:7}
\\
&
\left\langle \F_\gamma(t) \F_\gamma^+(t') \right\rangle= 
\kappa (\bar{n}_T+1) \delta(t-t'), \label{eq:8}
\\
&
\left\langle \F_\gamma^+(t) \F_\gamma^+(t') \right\rangle=\left\langle \F_\gamma(t) \F_\gamma(t') \right\rangle =0, \label{eq:9}
\end{align}
where $\bar{n}_T$ is the number of thermal photons in the cavity mode. We suppose that the temperature is low enough to set $\bar{n}_T=0$. In this case the correlations between the Langevin forces for a specific atomic ensemble are
\begin{align}
&\left\langle \F^\chi_a(t) \F^\chi_a(t') \right\rangle=\gamma_a\left\langle \N^\chi_a(t)\right\rangle \delta(t-t'), \label{eq:10}\\
&\left\langle \F^{\chi+}_M(t) \F^\chi_M(t') \right\rangle=\left(2 \gamma_{ab}-\gamma_a \right)
\left\langle \N^\chi_a(t)\right\rangle \delta(t-t'),
\label{eq:11}\\
&\left\langle \F^\chi_M(t) \F^{\chi+}_M(t') \right\rangle=
\Big(2 \gamma_{ab} \left\langle \N^\chi_b(t)\right\rangle + \gamma_a \left\langle \N^\chi_a(t)\right\rangle \Big)\nonumber \\
&\hphantom{\left\langle \F^\chi_M(t) \F^{\chi+}_M(t') \right\rangle=}
\times \delta(t-t'),\vphantom{\sum_i=1}\label{eq:12} \\
&\left\langle \F^{\chi+}_M(t) \F^{\chi+}_M(t') \right\rangle=
\left\langle \F^{\chi}_M(t) \F^{\chi}_M(t') \right\rangle = 0, \label{eq:13}\\
&\left\langle \F^{\chi}_a(t) \F^{\chi}_M(t') \right\rangle =
\left\langle \F^{\chi+}_M(t') \F^{\chi}_a(t) \right \rangle = 0   , \label{eq:14}\\
&\left\langle \F^{\chi}_a(t) \F^{\chi+}_M(t') \right\rangle= 
\gamma_a \langle \M^{\chi +}(t) \rangle  \delta(t-t') , \label{eq:15}\\
&\left\langle \F^{\chi}_M(t') \F^{\chi}_a(t)  \right\rangle= 
\gamma_a \langle \M^{\chi}(t) \rangle  \delta(t-t'). \label{eq:16}
\end{align}
Langevin forces corresponding to different atomic ensembles are uncorrelated.

%-----------Equivalent c-number stochastic Langevin equations equations-----
\subsection{Equivalent $c$-number stochastic Langevin equations}
\label{sec:model_cnumbers}
As a next step we introduce the stochastic $c$-number Langevin equations which are equivalent to the operator Langevin equations (\ref{eq:4}) and (\ref{eq:5}). To establish a unique relation between operators and $c$-number variables we have to define the correspondence between products of $c$ numbers and products of operators. In~\cite{Benkert90} and in a number of subsequent papers ~\cite{Yu08, Kolobov93, Davidovich96}, the normal ordering $\a^+, \, \M^+, \, \N_a, \, \N_b, \, \M, \, \a$ is chosen and the $c$-number Langevin equations were derived in such a way that the equations for first and second moments of $c$-number variables $\AA, \, \MM, \, \NN_a, \, \NN_b$ are identical to corresponding equations for normally ordered operator variables $\a, \, \M, \, \N_a, \, \N_b$. This choice leads to a redefinition of correlation functions for $c$-number Langevin forces $\FF_a, \, \FF_b, \, \FF_M, \, \FF_M*$. In our case this choice of normal ordering leads to self-contradicting correlation functions, see Appendix A for details. Moreover, the choice of normal ordering for atomic operators $\M, \N_a, \M^+$ is ambiguous, in contrast to the choice of normal ordering of field operators $\a$ and $\a^+$.

To overcome these difficulties we use here the following rules: (1) The equations for first moments of operators and $c$-number values should be identical; (2) the equations for second moments of $c$-number field variables $\AA$ and $\AA^*$ should be identical to the corresponding mean values of normally ordered field operators $\a$ and $\a^+$; (3) the equations for second moments of $c$-number atomic variables $\MM_\chi, \, \NN_a^\chi, \, \MM_{\chi}^*$ should be identical to corresponding equations for mean values of {\em symmetrized} products of operator variables $\M^\chi, \, \N_a^\chi, \M^{\chi+}$. Using these rules one can find the correlation functions for $c$-number Langevin forces in a similar way as in~\cite{Benkert90}:
\begin{align}
&
\left\langle \FF_a^\chi(t)\right\rangle=\left\langle \FF_b^\chi(t)\right\rangle=\left\langle \FF_M^\chi(t)\right\rangle=\left\langle \FF_\gamma(t)\right\rangle=0, \nonumber
\\
&
\left\langle \FF_\gamma^*(t) \FF_\gamma(t') \right\rangle = 
\left\langle \FF_\gamma^*(t) \FF_\gamma^*(t') \right\rangle  \label{eq:17} \\
&\hphantom{\left\langle \FF_\gamma^*(t) \FF_\gamma(t') \right\rangle} = \left\langle \FF_\gamma(t) \FF_\gamma(t')\right\rangle=0, \nonumber
\end{align}
and 
\begin{equation}
\left\langle \FF_k^\chi(t) \FF_l^\xi(t')\right\rangle=2 \DD^{\chi}_{kl}\delta(t-t')\delta_{\chi \xi}, \label{eq:18}
\end{equation}
where the diffusion coefficients $\DD^{\chi}_{kl}$ are
\begin{align}
&
2 \DD^\chi_{aa}=\gamma_a\left\langle \NN^\chi_a(t)\right\rangle, \label{eq:19}
\\
&
2 \DD^\chi_{\MM \MM}=2 \DD^\chi_{\MM^* \MM^*} = 0, \label{eq:20}
\\
&
2 \DD^\chi_{\MM \MM^*}=\gamma_{ab} \, \NN_\chi, \label{eq:21}
\\
&
2 \DD^\chi_{a \MM} = \frac{\gamma_a}{2}\MM_{\chi}, \label{eq:22}
\\
&
2 \DD^\chi_{a \MM^*} = \frac{\gamma_a}{2}\MM_{\chi}^*. \label{eq:23}
\end{align}
Here $\NN_\chi=\NN_a^\chi+\NN_b^\chi$. 

Equations (\ref{eq:4}) and (\ref{eq:5}) are already written in proper order, so that it is easy to get the equations for the corresponding $c$-number variables (for the sake of brevity, the argument $t$ is omitted here):
\begin{align}
&
\dot{\AA}=- 
\frac{\kappa}{2}\AA+ g \left(\MM_1 \Gamma_1+\MM_2 \Gamma_2 \right), 
\label{eq:24}
\\
&
\dot{\MM}_\chi=-\gamma_{ab} \MM_\chi + g \Gamma_\chi \left(2 \NN^\chi_a-\NN_\chi \right) \AA +\FF_M^\chi, \label{eq:25}
\\
&
\dot{\NN}_a^\chi=-\gamma_{a} \NN^\chi_a - g \Gamma_\chi 
\left(\AA^*\MM_\chi+ \AA\MM_{\chi}^* \right) +\FF_a^\chi. \label{eq:26}
\end{align}

In the bad-cavity regime, the field variable $\AA$ follows the atomic variables adiabatically:
\begin{equation}
\AA=\frac{2g}{\kappa}\left(\MM_1 \Gamma_1+\MM_2 \Gamma_2\right). \label{eq:27}
\end{equation}
It yields the full set of equations for the atomic variables
%-------------------------------------------------------------------------------
%
\begin{widetext}
\begin{align}
& \dot{\MM_1}(t)=-\gamma_{ab}\MM_1(t)+
G\Big[\MM_1(t) \Gamma_1^2(t)+\MM_2(t)  \Gamma_1(t) \Gamma_2(t) \Big] 
\left(\NN_a^1(t)-\frac{\NN_1}{2}\right)+\FF_M^1(t), \label{eq:28}
\\
&
\dot{\NN}_a^1(t)=-\gamma_{a}\NN_a^1(t)-
G\left[\MM_1(t) \MM_1^*(t) \Gamma_1^2(t)+\frac{\MM_1(t) \MM_2^*(t)+\MM_1^*(t) \MM_2(t)}{2}\Gamma_1(t) \Gamma_2(t) \right]+\FF_a^1(t), \label{eq:29}
\\
& \dot{\MM_2}(t)=-\gamma_{ab}\MM_2(t)+
G\Big[\MM_2(t) \Gamma_2^2+\MM_1(t)  \Gamma_1(t) \Gamma_2(t) \Big] 
\left(\NN_a^2(t)-\frac{\NN_2(t)}{2}\right)+\FF_M^2(t), \label{eq:30}
\\
&
\dot{\NN}_a^2(t)=-\gamma_{a}\NN_a^2(t)-
G\left[\MM_2(t) \MM_2^*(t) \Gamma_2^2(t)+\frac{\MM_1(t) \MM_2^*(t)+\MM_1^*(t) \MM_2(t)}{2}\Gamma_1(t) \Gamma_2(t) \right]+\FF_a^2(t), \label{eq:31}
\end{align}
\end{widetext}
%
%-------------------------------------------------------------------------------
where $G=4g^2/\kappa$. Using Eqs. (\ref{eq:28})--(\ref{eq:31}) and correlation functions for Langevin forces (\ref{eq:19})--(\ref{eq:23}), we simulate the lasing process.
%
%-----------Simulation of c-number Langevin forces------------------------------
\subsection{Simulation of $c$-number Langevin forces}
\label{sec:model_cnumbers}
In the present work we solve the set of equations (\ref{eq:28})--(\ref{eq:31}) using the simplest Euler algorithm. Namely, introducing vectors $x=\left(\MM_1, \N_a^1, \MM_2, \N_a^2\right)$ we rewrite (\ref{eq:28})--(\ref{eq:31}) as
\begin{equation}
\dot{x}(t)=r(t,x)+f(t),
\label{eq:32}
\end{equation}
where $r(t,x)$ is the deterministic (``regular'') part of the right-hand side of Eqs. (\ref{eq:28})--(\ref{eq:30}), and $f(t)=\left(\FF_M^1(t), \FF_a^1(t), \FF_M^2(t), \FF_a^2(t) \right)$ is a stochastic part. Then the algorithm is
\begin{equation}
x(t_{i+1})=x(t_i)+\left[ r(t_i,x)+\tilde{f}(t_i) \right]\Delta t,
\label{eq:33}
\end{equation}
where $\Delta t=t_{i+1}-t_i$, and 
\begin{equation}
\tilde{f}(t_i)=\frac{1}{\Delta t}\int_{t=t_i}^{t_{i+1}}f(t)dt \label{eq:34}
\end{equation}
is a set of random variables with zero mean and with correlations
\begin{equation}
\left\langle \tilde{f}_k(t_i)\tilde{f}_m(t_j)\right\rangle = \frac{2\DD_{km} \delta_{ij}}{\Delta t}. \label{eq:35}
\end{equation}
Here $\DD_{km}$ is one of the diffusion coefficients (\ref{eq:19})--(\ref{eq:23}). Note that stochastic terms corresponding to different atomic ensembles are independent.

It is convenient to introduce two real column vectors $\FFF^1(t_i)$ and $\FFF^2(t_i)$ such that
\begin{equation}
\FFF^\chi(t_i)=\left(
\begin{array}{c}
\tilde{\FF}_a^\chi (t)\\
\mathrm{Re} \big[\tilde{\FF}_M^\chi(t_i)\big]\vphantom{\Bigg{()}}\\
\mathrm{Im} \big[\tilde{\FF}_M^\chi(t_i)\big] 
\end{array}
\right).
\label{eq:36}
\end{equation}
Then the correlators (\ref{eq:35}) can be rewritten in matrix form as
\begin{equation}
\left\langle\FFF^\chi(t_i)(\FFF^\chi(t_j))^T\right\rangle=\frac{\delta_{ij}}{\Delta t}\DDD^{\chi},
\label{eq:37}
\end{equation}
where the covariance matrix $\DDD^{\chi}$ can be derived from (\ref{eq:19})--(\ref{eq:23}) and (\ref{eq:35}):
%
%\newcolumntype{C}{>{\displaystyle}c}
\begin{equation}
\DDD= \frac{1}{\Delta t}\left(
\begin{array}{ccc}
\gamma_a \NN_a & \mathrm{Re}[\MM]\displaystyle{\frac{\gamma_a}{2}} & \mathrm{Im}[\MM]\displaystyle{\frac{\gamma_a}{2}} \\
\mathrm{Re}[\MM]\displaystyle{\frac{\gamma_a}{2}} & \NN \displaystyle{\frac{\gamma_{ab}}{2}} & 0 \vphantom{\Bigg(} \\
\mathrm{Im}[\MM]\displaystyle{\frac{\gamma_a}{2}} & 0 &\NN \displaystyle{\frac{\gamma_{ab}}{2}}
\end{array}
\right).
\label{eq:38}
\end{equation}
(Here and below in this section an index $\chi$ and an argument $t_i$ are omitted for the sake of brevity.) The matrix $\DDD$ is positive semi-definite according the Sylvester's criterion. Indeed, the determinant of this matrix is
\begin{equation} 
\left| \DDD \right|=\frac{\NN \gamma_{ab} \gamma_a}{4 \Delta t} 
\left[ \NN \NN_a \gamma_{ab}-\MM \MM^*\frac{\gamma_a}{2} \right]\geq 0 \label{eq:39}.
\end{equation} 
because $\left|\MM\right|^2\leq\NN_a \NN_b\leq\NN_a \NN$, and $\gamma_{ab} \geq {\gamma_a/2}$.

Consider a column vector 
\begin{equation}
\AAA=\XXX^{-1} \cdot \FFF, \label{eq:40}
\end{equation}
where $\XXX=(X^{(1)},X^{(2)},X^{(3)})$ is a matrix constructed of the normalized eigenvectors of $\DDD$ corresponding to the eigenvalues $\lambda_1$,  $\lambda_2$, $\lambda_3$. Correlations of elements of $\AAA$ can be found as
\begin{equation}
\left\langle\AAA_k(t_i)\AAA_m(t_j)\right\rangle=\delta_{ij} \delta_{km} \Lambda_m.
\label{eq:41}
\end{equation}
Therefore the elements of $\AAA$ are independent random values with zero means and given dispersions. These values can be easily simulated, and the Langevin forces can then be calculated by the inversion of (\ref{eq:40}).

%-------------------------------------------------Simulation---------------------------------------------------------
\section{Simulation}
\label{sec:sim}
\begin{figure}
\begin{center}
\resizebox{0.48\textwidth}{!}{\includegraphics{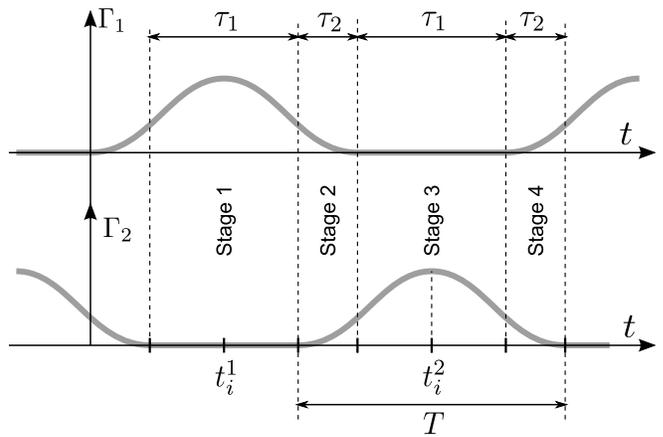}}
\end{center}
\caption{Time-dependence of atom-cavity coupling expressed by the coupling functions $\Gamma_1(t)$ and $\Gamma_2(t)$ for different stages of the lasing cycle.}
\label{fig:f2}
\end{figure}

In this section we specify our model and present the results of simulations of the lasing process.
Consider the functions $\Gamma_1(t)$ and $\Gamma_2(t)$ describing the time dependence of the coupling between an atomic ensemble and the cavity mode. We suppose that this coupling is provided by the motion of the optical lattice confining the atoms. It seems to be necessary to keep a constant velocity of the atomic ensemble during its motion through the cavity to avoid the time-dependent second-order Doppler effect. Usually the transversal distribution of the electric field is Gaussian. For this reason, we take the functions $\Gamma_1(t)$ and $\Gamma_2(t)$ of the form
\begin{equation}
\Gamma_{\chi}(t)=\exp\left[-\frac{18\, (t-t^\chi_i)^2}{T^2} \right],
\label{eq:42}
\end{equation}
in the $i$th lasing cycle. Here $t^\chi_i$ is the time when the $\chi$th ensemble crosses the symmetry axis of the cavity, and $T$ is the time for moving the atomic ensemble through the cavity between two points where the amplitude of the intracavity field is $e^{-9/2}\simeq 1 \%$ of its value (on the cavity axis). We neglect any coupling outside these points. Therefore $T=\tau_1+2 \tau_2$, where $\tau_1$ is the duration of stages 1 and 3, and $\tau_2$ is the duration of stages 2 and 4; see Fig.~\ref{fig:f2}.
%---------------------------------begin figure ---------------------------
%---------------------------------end figure-----------------------------

We now simulate the lasing process and study the dependence of intracavity field phase fluctuations on the temporal parameters of the lasing cycle, summarized in the ratio
\begin{equation}
R=\frac{\tau_1}{\tau_1+\tau_2},
\label{eq:43}
\end{equation}
which describes the temporal overlap of different atomic ensembles inside the cavity.
It follows from general considerations, that the lower the $R$ (more overlap), the better the polarization of the ``old" ensemble will be inherited by the ``fresh'' one. An ideal system would be a continuous delivery of inverted atoms trapped in the optical conveyor belt, however such a device remains to be realized experimentally. If the experimental setup traps the atoms in a pulsed regime and if it contains only two moving optical lattices, then the atomic ensemble should be loaded into the optical lattice, prepared in the upper lasing state, and transferred into the cavity during the time $\tau_1$. Therefore $\tau_1$ (and $R$) can not be made arbitrarily small. This difficulty could in principle be overcome by scaling to more than two moving optical lattices in one setup, however, significantly enhancing the experimental complexity. Therefore it is interesting to find the optimal range of $R$ where the phase transfer between two ensembles still works.

In our simulation we use the following parameters for the atomic ensemble and the laser cavity: $\kappa=10^{5}$ s$^{-1}$ (this value can be attained in a cavity with a finesse of about $5\times 10^4$  and a cavity length about 10\,cm), $\gamma_a=\gamma_{ab}=5 \times 10^{-3}$ s$^{-1}$ (this value is typical for the spontaneous decay rate of the $^3P_2$ state of divalent atoms, see \cite{Porsev04, Yasuda04}), $g=2.5$ s$^{-1}$ (this value corresponds to a cavity mode waist of about 1\,mm, the wavelength of the lasing transition is about 700\,nm and the cavity length is about 10\,cm), and an average number of atoms in one ensemble $\overline{\NN}=2\times 10^5$. An important feature of any realistic preparation and trapping of the atomic ensembles is that the number of trapped atoms fluctuates. We suppose that the dispersion of the number of atoms in one ensemble is 10\,\%.

%---------------------------------begin figure ---------------------------
\begin{figure}
\begin{center}
\resizebox{0.48\textwidth}{!}{\includegraphics{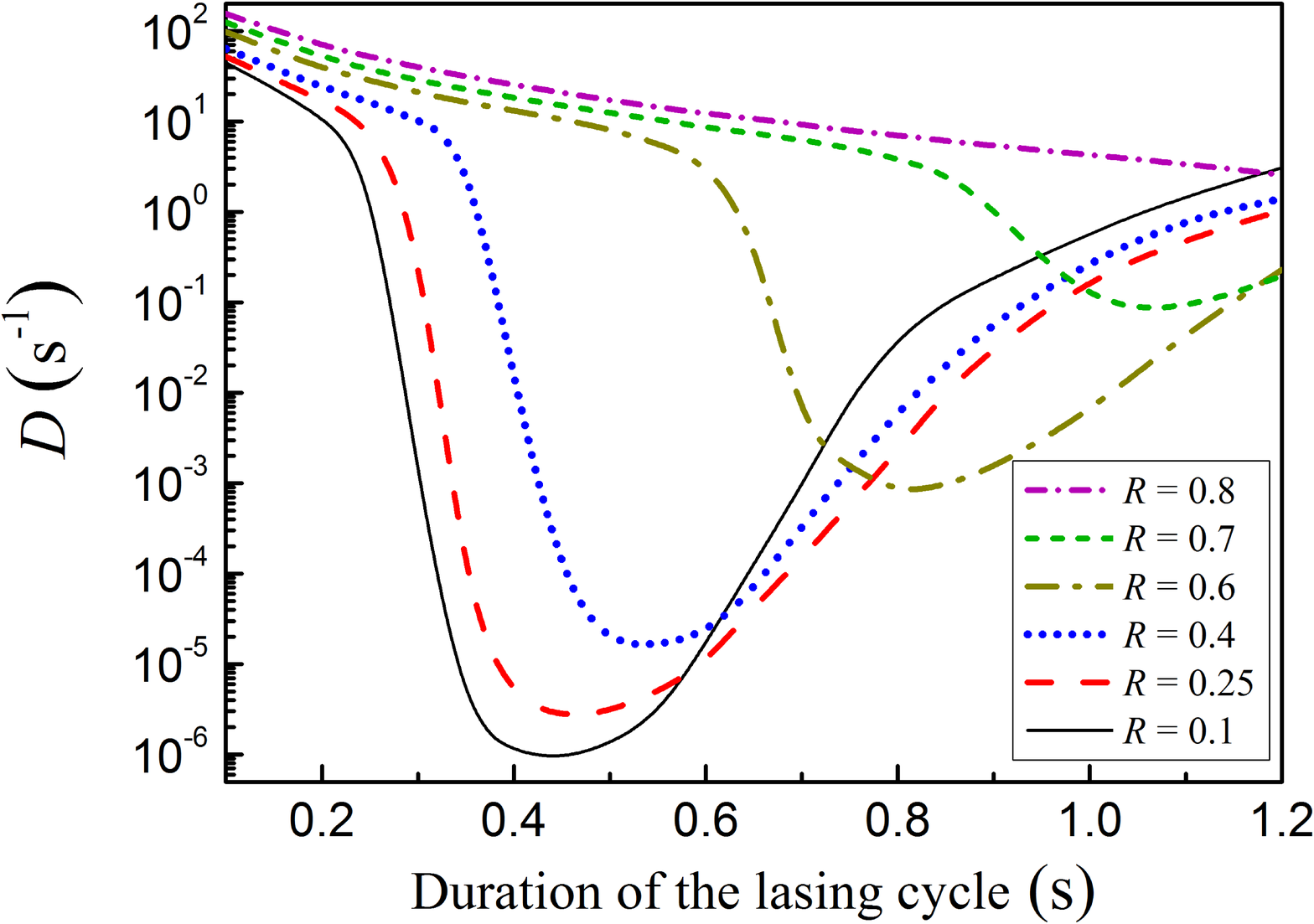}}
\end{center}
\caption{(Color online) Phase diffusion coefficient versus the duration of the lasing cycle for different values of the ensemble overlap parameter $R$.}
\label{fig:f3}
\end{figure}
%---------------------------------end figure-----------------------------

For all lasing stages we simulate the lasing process according the method described in Sec.~\ref{sec:model} using the functions (\ref{eq:42}). At the start of stage 2 and stage 4 of the lasing process, we chose the number $\NN$ of atoms in the second and the first ensemble respectively, including random fluctuations. We suppose that atoms enter the cavity mode in the upper lasing state $|a\rangle$ with zero polarization. We then trace the intracavity field variable $\AA$ over 1500 lasing cycles. This process is repeated several times which allows us to study the Allan dispersion $\sigma_\phi^2$ of the phase $\phi$ of the laser field:
\begin{equation}
\sigma_\phi^2 (\Delta t)=\frac{1}{2}\left\langle (\phi_i-\phi_{i+1})^2 \right\rangle,
\label{eq:44}
\end{equation}
where $\phi_i$ and $\phi_{i+1}$ are the phases averaged over the $i$th and $(i+1)$th adjanced time intervals, each of which has a duration of $\Delta t$. In our case $\sigma_\phi^2 (\Delta t) \propto \Delta t$, as is typical for a random phase walk. This process can be approximately described by the equation
\begin{equation}
\frac{d\phi}{dt}=f(t), \, \, \langle f(t) \rangle=0, \, \, \langle f(t) f(t') \rangle=2 D \delta(t-t').
\label{eq:45}
\end{equation}

It is easy to show that the phase diffusion coefficient $D$ is connected with the Allan dispersion $\sigma_\phi^2$ as
\begin{equation}
\sigma_\phi^2 (\Delta t) \simeq \frac{2 D}{3} \Delta t.
\label{eq:46}
\end{equation}
So one can easily find the phase diffusion coefficient $D$ using the calculated Allan dispersion.

%---------------------------------begin figure ---------------------------
\begin{figure}
\begin{center}
\resizebox{0.48\textwidth}{!}{\includegraphics{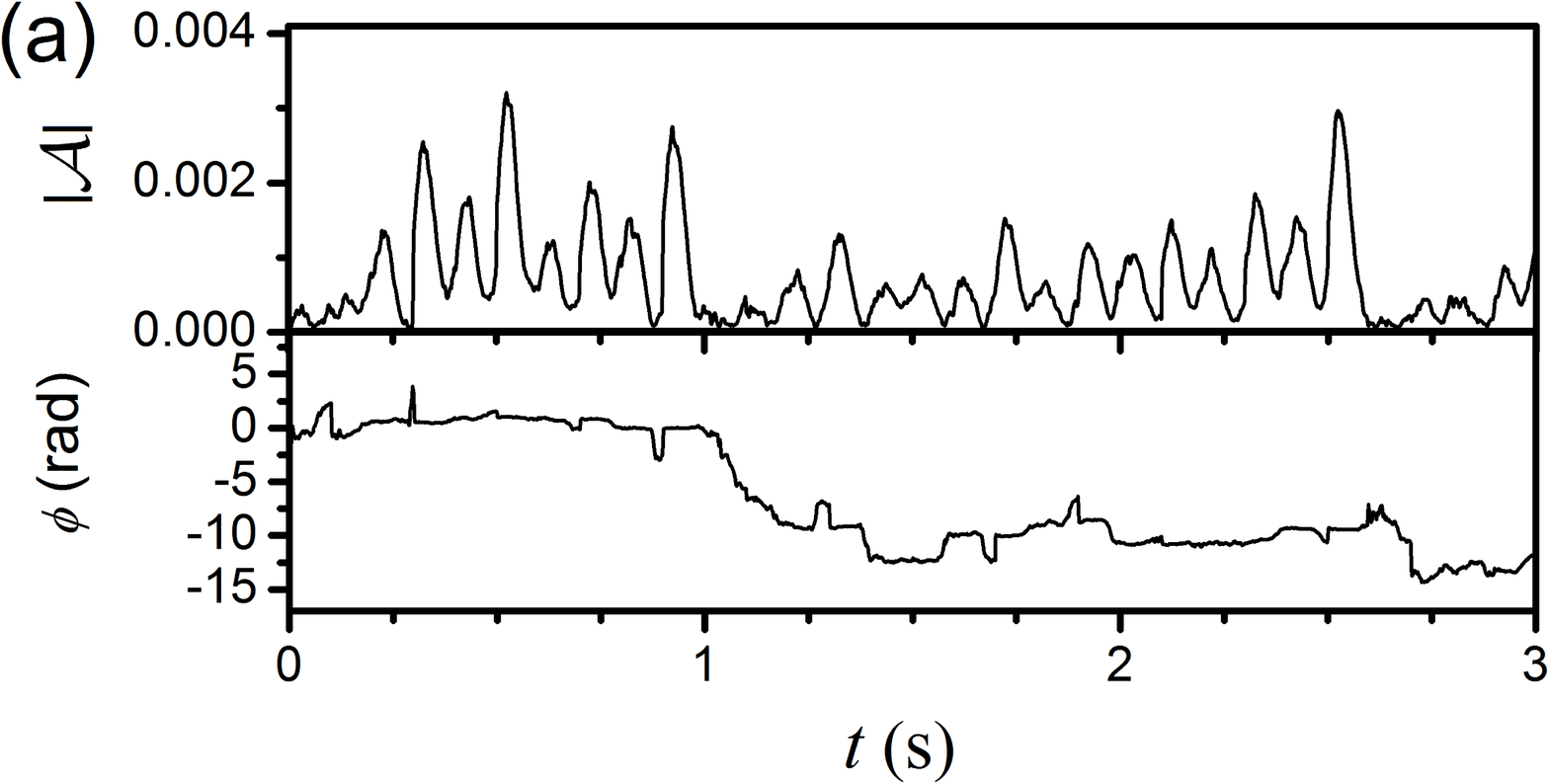}}\\
\resizebox{0.48\textwidth}{!}{\includegraphics{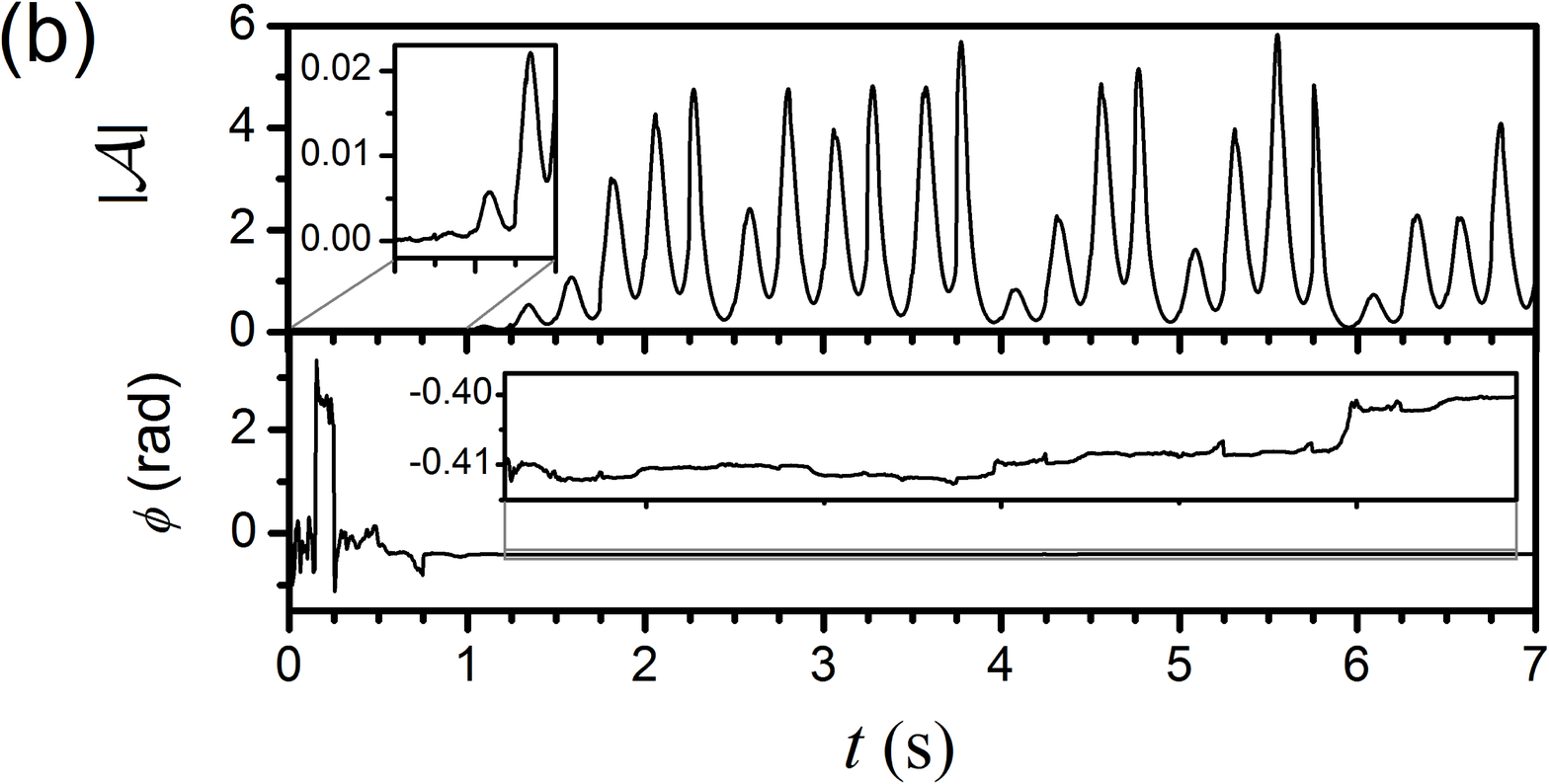}}\\
\resizebox{0.48\textwidth}{!}{\includegraphics{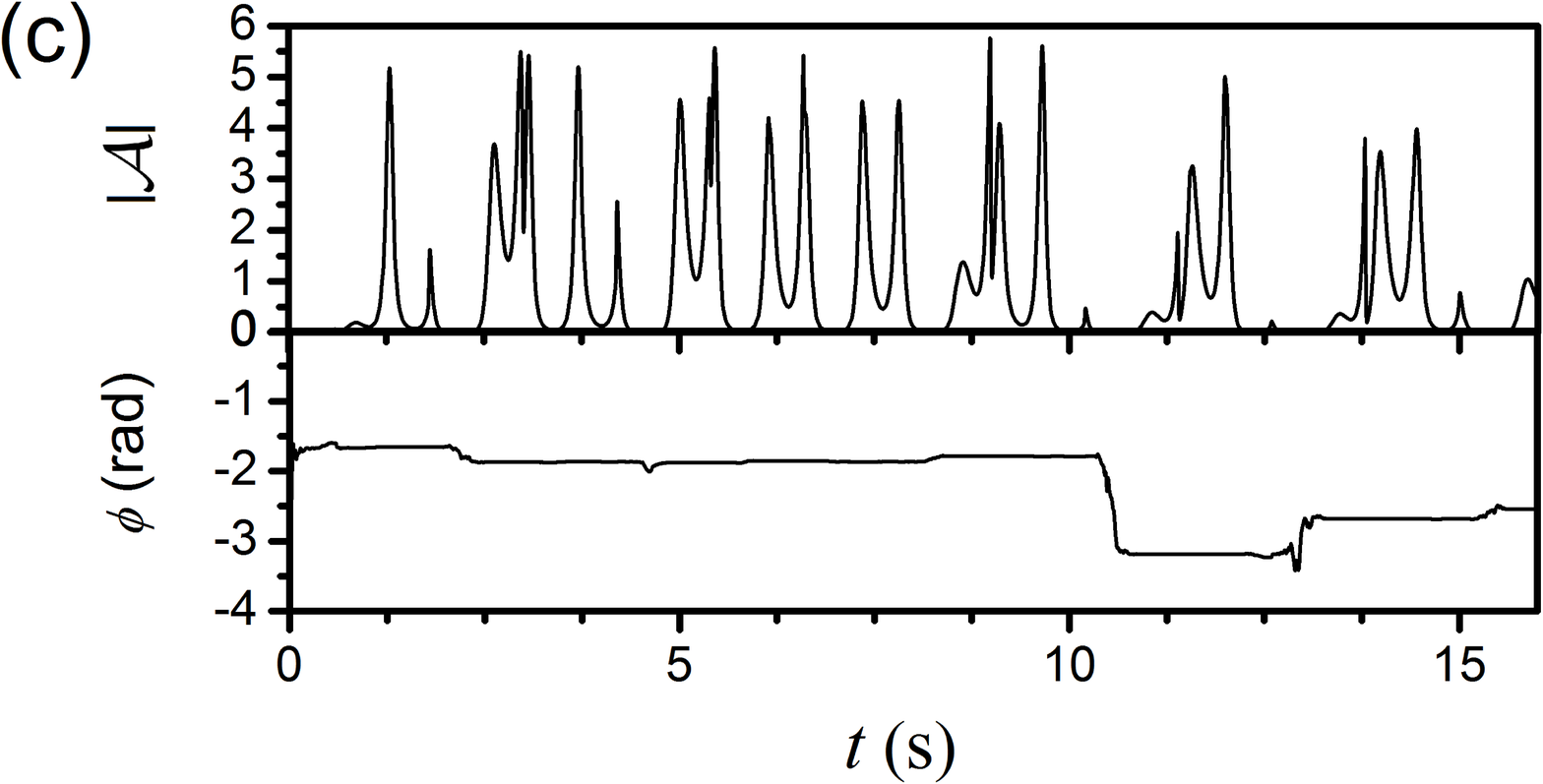}}
\end{center}
\caption{Behavior of amplitude and phase of the intracavity field in a particular realization over several lasing cycles for $R=0.25$ and different durations of lasing cycle. (a) $2 (\tau_1+\tau_2)=0.2 \, \mathrm{s}$; (b) $ 2(\tau_1+\tau_2)=0.5 \, \mathrm{s}$, behaviour of the amplitude in the first two lasing cycles and of the phase in established regime are shown in the insets; (c) $2 (\tau_1+\tau_2)=1.2 \, \mathrm{s}$.}
\label{fig:f4}
\end{figure}
\begin{figure}[t]
\begin{center}
\resizebox{0.48\textwidth}{!}{\includegraphics{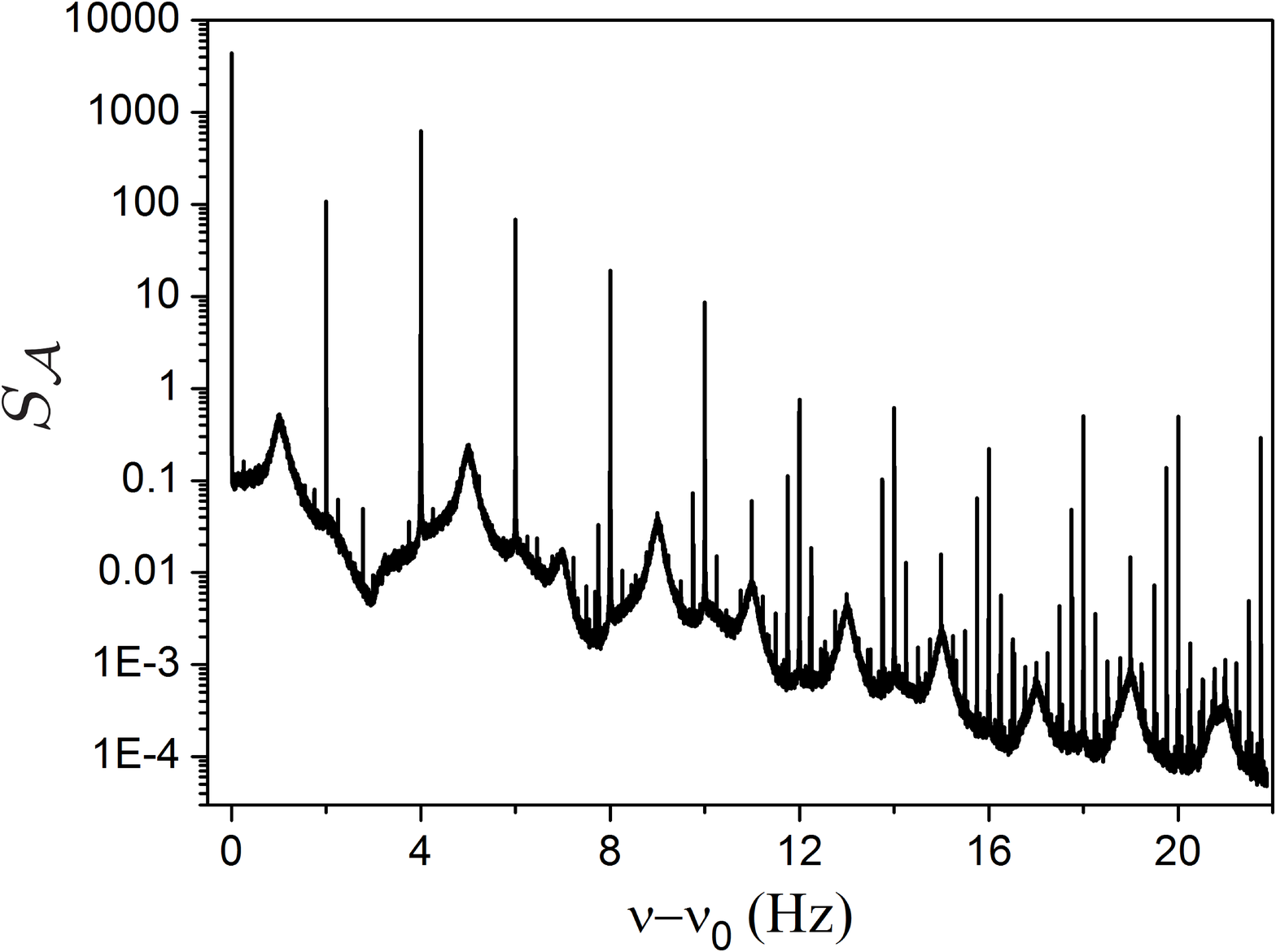}}
\end{center}
\caption{(Color online) Power spectral density of the intracavity field for $\tau_1=0.0625$ s, $\tau_2=0.1775$ s.}
\label{fig:f5}
\end{figure}

%---------------------------------end figure-----------------------------
The phase diffusion coefficient is of significance here, as for laser radiation with negligible amplitude fluctuations and with phase fluctuations described by a Markovian process (\ref{eq:45}), the power spectral density is a Lorentzian function with linewidth at half maximum $\Delta \omega=2D$~\cite{Cook80}. In our case, the laser operates in the pulsed regime and the spectrum has a more complicated shape, as indicated in Fig.~\ref{fig:f5}. Here, $2D$ is the linewidth of the narrow central peak corresponding to the main carrier component of the spectrum.

Dependencies of the phase diffusion coefficient $D$ on the duration $2 (\tau_1+\tau_2)$ of the lasing cycle for different values of $R$ are presented in Fig.~\ref{fig:f3}. To discuss these results, let us consider the time dependencies of amplitude $|\AA|$ and phase $\phi$ of the intracavity field $\AA$ for some particular realizations; see Fig.~\ref{fig:f4}. We see that if the lasing cycle is too short, the atoms have not enough time to develop a significant polarization. Therefore almost all the atoms leave the cavity before they can transfer their energy to the cavity mode, and the intracavity field is extremely weak; see Fig.~\ref{fig:f4}(a). Weakness of the intracavity field leads to a comparatively large contribution of Langevin forces which in turn leads to a large phase diffusion coefficient. On the other hand, if the lasing cycle is too long, the atoms of the first ensemble transfer all their energy to the intracavity field during stage 1, see Fig.~\ref{fig:f4} (c). Therefore at the beginning of stage 2, when the ``fresh'' ensemble is introduced into the cavity, the first ensemble has already lost the polarization. Of course, any transfer of polarization and passing on of the phase is impossible in this case. 

For the optimal duration of the lasing cycle, atoms have enough time to develop significant polarization, and this polarization survives until the introduction of the next atomic ensemble into the cavity, see Fig.~\ref{fig:f4} (b). In this case, during the first few lasing cycles, the amplitude of the pulses grows from pulse to pulse [see inset in the upper graph in Fig.~\ref{fig:f4} (b)], and the phase fluctuates considerably [see the lower graph in Fig.~\ref{fig:f4} (b)]. After several cycles, a quasi stationary regime with phase-linking between successive ensembles is established, and only weak phase diffusion remains [see inset in the lower graph in Fig.~\ref{fig:f4} (b)]. Note that for non optimal regimes presented in Fig.~\ref{fig:f4}(a) and ~\ref{fig:f4}(c), the phase fluctuations  in the first few pulses are practically the same as in the quasi stationary regime; the phase can not be kept longer than a few cycles.

We obtain optimal values for the lasing cycle of about $0.4$--$0.6$ s length for the experimental parameters introduced above. This result can be understood intuitively by comparing it to the characteristic duration $t_p$ of a single superradiance pulse of an atomic ensemble coupled to the cavity, see Appendix B. For a single ensemble ($\NN=2\times 10^5$, $G=2.5 \times 10^{-4}$) we estimate $t_p\simeq 0.1$ s. In the dynamic implementation discussed here, this time should be increased following the time dependence of $\Gamma(t)$. Because one lasing cycle consists of two superradiation pulses, one understands that the optimal timing of these pulses corresponds to about 2--3\,$t_p$. Also we note that the optimal lasing cycle duration decreases with smaller $R$ because the simultaneous coupling of two atomic ensembles with the cavity mode results in a shorter radiation pulse.

%---------------------------------begin figure ---------------------------
%---------------------------------end figure-----------------------------

Finally, let us consider the one-sided power spectral density
\begin{align}
S_{\AA}(\nu)&=\frac{\left|\AA_{\omega}\big(2 \pi (\nu-\nu_0)\big)\right|^2}{\pi} \nonumber \\
&=4\int\limits_0^\infty \left\langle
 \mathrm{Re}\, \big(\AA (0) \AA^*(\tau)\big)
 \right\rangle \cos (2\pi \nu) d\tau,  
\end{align}
where $\nu_0=2\pi\omega_{ab}$. The power spectral density for $\tau_1+\tau_2=0.25 \, \mathrm{s}$ (which corresponds to a total duration of one lasing cycle of 0.5\,s), $R=0.25$, averaged over 60 realizations, each with 1500 cycles, is represented in Fig.~\ref{fig:f5}. One can observe a high central peak corresponding to the main lasing component with the frequency $\nu_0$, and a number of sufficiently lower sideband peaks at frequencies corresponding to multiples of the repetition frequency of the transport cycles (2 Hz in our case).

%------------------------------------------------Conclusion---------------------------------------------------------
\section{Conclusion}
\label{sec:concl}
We proposed to keep the phase coherence between pulses of a bad-cavity active optical frequency standard by sequential coupling and decoupling of prepumped atomic ensembles with the optical cavity mode. We studied the phase diffusion of the intracavity field and demonstrated that for the optimal duration of the lasing cycle (about a few durations of the superradiation pulse of a single atomic ensemble), a phase diffusion coefficient $D$ as low as a few $10^{-6}$ can be realized. We found that this phase relay process is robust to 10\,\%  fluctuations of the atom number introduced into the cavity. We conclude that the method of sequential coupling may be a promising approach towards the creation of an active optical frequency standard.

%Acknowledgments---------------------------------------------------------

\section{Acknowledgments}
We thanks M. Kolobov and N. Larionov for helpful discussion. This research was supported by the Austrian Science Fund (FWF) through Project No. M1272-N16 (TheoNAC).

%-------------------------------------------------Appendix---------------------------------------------------------
\section*{APPENDIX A: SELF-CONTRADICTION OF CONVENTIONAL CHOICE OF CORRELATION FUNCTIONS}

%%%%%%%%%%%%%%%%%%%%%%%%%%%%%%%%%%%%%%%%%%%%%%%%%%%%%%%%%%%%%%%%%%%%%%%%%%%%%%%%%%%%%%%%%%%%%
%\subsection{{\em c}-number Langevin forces: self-contradiction of conventional choice of correlation functions}
\label{sec:append}
As it was mentioned in Sec.\,\ref{sec:model_cnumbers}, usually the correspondence between operator and $c$-number variables is based on the requirement that products of $c$-number variables correspond to the normally ordered operator variables with the normal ordering $\a^+, \, \M^+, \, \N_a, \, \N_b, \, \M, \, \a$ \cite{Benkert90}. This convention gives the diffusion coefficients $\DD_{kl}^{\chi}$, different from (\ref{eq:19})--(\ref{eq:23}). Instead, one can obtain (see \cite{Kolobov93} for details)
\begin{align}
&
2 \DD^\chi_{aa}=\gamma_a\left\langle \NN^\chi_a(t)\right\rangle  \nonumber \\
&\hphantom{2 \DD^\chi_{aa}=} -g \Gamma_\chi(t) 
\left\langle \MM_\chi^*(t) \AA(t) + \MM_\chi(t) \AA^*(t) \right\rangle, \label{eq:a1}
\\
&
2 \DD^\chi_{\MM \MM}=2 g \Gamma_\chi (t) \left\langle \MM_\chi(t) \AA(t) \right\rangle,
\vphantom{\Big(} \label{eq:a2}
\\
&
2 \DD^\chi_{\MM^* \MM^*}=2 g \Gamma_\chi (t) \left\langle \MM_\chi^*(t) \AA^*(t)\right\rangle, \label{eq:a3}
\\
&
2 \DD^\chi_{\MM \MM^*}=(2\gamma_{ab}-\gamma_a)\left\langle \NN_a(t)\right\rangle, \vphantom{\Big(} \label{eq:a4}
\\
&
2 \DD^\chi_{a \MM} = 2 \DD^\chi_{a \MM^*}=0. \label{eq:a5}
\end{align}

If one tries to use these diffusion coefficients for the numerical simulation of Langevin forces, one obtains an inconsistency. Indeed, if one introduces the real column vector (\ref{eq:36}), then the covariance matrix $\DDD$ occurs to be
\begin{widetext}
\begin{equation}
\DDD = \frac{1}{\Delta t}\left(
\begin{array}{ccc}
\gamma_a \NN_a - G \MM \MM^* & 0 & 0 \\
0 &\hphantom{aaaa}  \displaystyle{\NN_a \left(
\gamma_{ab}-\frac{\gamma_{a}}{2}\right) + G \Gamma^2 \frac{\MM^2+\MM^{*2}}{4} 
\hphantom{aaaa} \vphantom{\frac{\Big{(}}{\Big{(}}}} 
& 
\displaystyle{G \Gamma^2 \frac{\MM^2-\MM^{*2}}{4 i} }
\\
0 & \displaystyle{G \Gamma^2 \frac{\MM^2-\MM^{*2}}{4 i} }
&
\displaystyle{\NN_a \left(
\gamma_{ab}-\frac{\gamma_{a}}{2}\right) - G \Gamma^2 \frac{\MM^2+\MM^{*2}}{4} }
\end{array}
\right)
\label{eq:a6}
\end{equation} 
\end{widetext}
after adiabatic elimination of the field variables $\AA$, $\AA^*$. This matrix must be positive-semidefinite because it is a covariance matrix of real random values $\FF_a$, $\mathrm{Re}[\FF_\MM]$, $\mathrm{Im}[\FF_\MM]$. However, this is not true in the general case. For example, if $\gamma_{ab}=\gamma_a/2$, the last diagonal element of the matrix $\DDD$ is always negative. The first diagonal element is also negative, if $G \gg \gamma_a$ (which corresponds to the high-cooperativity regime) and if the atomic polarization is not drastically suppressed by some reasons. In any of these cases, the matrix $\DDD$ is not positive-semidefinite according to Sylvester's criterion. So the conventional choice of correlation functions occurs to be self-contradicting.

%%%%%%%%%%%%%%%%%%%%%%%%%%%%%%%%%%%%%%%%%%%%%%%%%%%%%%%%%%%%%%%%%%%%%%%%%%%%%%%%%%%%%%%%%%%%%
\section*{APPENDIX B: DURATION AND SHAPE OF SUPERRADIANCE IMPULSE}
\label{sec:append_single} 
Consider one single ensemble of $\NN$ two-level atoms coupled to the resonance mode of the bad cavity in the high-cooperativity regime, i.e. when $\NN g^2 \gg \gamma_{ab} \kappa$. We study the time dependence of the intracavity field using the set of equations for $c$-number variables $\AA$, $\MM$, $\NN_a$ when $\Gamma(t)=1$. For the sake of simplicity we neglect the Langevin forces and suppose that initially all the atoms are in the upper lasing state. Of course such simplification obstructs the correct behavior {\em when} the superradiance pulse starts. However, this model can reproduce the shape of the pulse because the interaction of the atomic ensemble with the thermal bath will play a significant role only in the very beginning, or in the very end of the pulse.

We adiabatically express the field variable $\AA=2g\MM/\kappa$ via the atomic polarization $\MM$ similarly to (\ref{eq:27}). Then the set of equations for the atomic variables reads as
\begin{eqnarray}
\dot{\MM}&=&-\gamma_{ab}\MM+G \MM \left(\NN_a-\frac{\NN}{2} \right) \label{eq:a7} \\ 
\dot{\NN}_a&=&-\gamma_{a}\NN_a-G \MM \MM^*, \label{eq:a8}
\end{eqnarray} 
where $G=4g^2/\kappa$. In the high-cooperativity regime we neglect the first terms on the right-hand side of these equations. Then we introduce $Q=\MM \MM^*$. The set of equations transforms to
\begin{eqnarray}
\dot{Q}&=&G(\NN-2\NN_a)Q \label{eq:a9} \\ 
\dot{\NN}_a&=&-G Q. \label{eq:a10}
\end{eqnarray} 
The solution to this set of equations reads as
\begin{eqnarray}
Q(t)&=&\frac{\NN^2}{4 \cosh^2[G \NN (t-t_0)/2]} \label{eq:a11} \\ 
\NN_a(t)&=&\frac{\NN}{1+\exp[G \NN (t-t_0)]}, \label{eq:a12}
\end{eqnarray} 
where $t_0$ is the time when the atomic polarization (and therefore the intracavity field) is maximal. Here we took into account that $\NN_a(-\infty)=\NN$, $\MM(-\infty)=0$.

The superradiance pulse has the shape
\begin{equation}
\AA(t)=\frac{2g}{\kappa}\frac{\NN \exp(i \phi)}{2 \cosh[G \NN (t-t_0)/2]}, \label{eq:a13}
\end{equation}
where $\phi$ is the phase of the intracavity field. The characteristic duration $t_p$ of this pulse (measured to the half of its height) is 
\begin{equation}
t_p = \frac{4 \mathrm{arccosh \,} 2}{G \NN} \simeq \frac{5.27}{GN}. \label{eq:a14}
\end{equation}
%

%-------------------------------------------------Bibliography--------------------------------------------------------------

\renewcommand{\refname}{Literature}

\end{document}